\documentclass[11pt,a4paper]{article}
\usepackage{amsmath,amssymb}
\usepackage{epsfig,graphicx}

\newcommand{\ov}{\overline}
\newcommand{\nn}{\nonumber}
\newcommand{\be}[1]{\begin{equation}\label{#1}}
\newcommand{\ee}{\end{equation}}
\newcommand{\ba}[1]{\begin{eqnarray}\label{#1}}
\newcommand{\ea}{\end{eqnarray}}
\newcommand{\rf}[1]{(\ref{#1})}

\def\stackunder#1#2{\mathrel{\mathop{#2}\limits_{#1}}}

\begin{document}

\title{\bf Conventional cosmology from multidimensional models}
\author{A.~I.~Zhuk\footnote{{\bf e-mail}:
zhuk@paco.net}
\\
\small{\em Department of Theoretical Physics and Astronomical
Observatory,}\\
\small{\em Odessa National University,} \\
\small{\em 2 Dvoryanskaya St., Odessa 65026, Ukraine} }
\date{}
\maketitle

\begin{abstract}
We investigate a possibility for construction of the conventional
Friedmann cosmology for our observable Universe if underlying
theory is multidimensional Kaluza-Klein model endowed with a
perfect fluid. We show that effective Friedmann model obtained by
dynamical compactification of the multidimensional one is faced
with too strong variations of the fundamental "constants". From
other hand, models with stable compactification of the internal
space are free from this problem and also result in conventional
4D cosmological behavior for our Universe. We prove a no-go
theorem which shows that stable compactification of the internal
spaces is possible only if equations of state in the external and
internal spaces are properly adjusted to each other. With a proper
choice of parameters (fine tuning), effective cosmological
constant in this model provides the late time acceleration of the
Universe. The fine tuning problem is resolved in the case of the
internal spaces in the form of orbifolds with branes in fixed
points. However, in this case the effective potential is too flat
(mass gravexcitons is very small) to provide necessary constancy
of the effective fundamental "constants".
\end{abstract}


\section{Introduction}

\setcounter{equation}{0}

Multidimensionality of our Universe is one of the most intriguing
assumption in modern physics. It is a natural ingredient of
theories which unify different fundamental interactions with
gravity, such like string/M-theory. It is of great interest to
investigate the cosmological consequences of this assumption. One
of the most simple and natural generalization leads to the
Kaluza-Klein (KK) models with warped product topology
\be{0.0}
M = M_0\times M_{D'} \ee
consisting of an external ("our") four-dimensional spacetime
manifold $M_0$ and a $D'$-dimensional compact space component
$M_{D'}$. Here, dynamics of factor spaces is described by their
own scale factors. Obviously, dynamical picture of this model can
be very complicated and significantly differ from the evolution of
the Friedmann-Robertson-Walker (FRW) Universe. This deviation may
have dramatic consequences. For example, the abundance of the
light elements is very sensitive to the rate of the evolution.
Thus, the deviations from evolution of the FRW Universe may
contradict observable data. Hence, it is very important to show
that proposed new cosmological models are in concordance with
conventional FRW cosmology (at least during the radiation and
matter dominated stages). It is also very desirable within these
models to have stages of the initial inflation as well as late
time acceleration. The main purpose of the present paper consists
in investigation of a possibility of the conventional description
of effective four-dimensional cosmological models obtained from
Kaluza-Klein models.

We investigate KK models where spacetime  is endowed with a
multicomponent perfect fluid \cite{CQG(1996)}. Within the standard
KK models and according to the present level of the experimental
data, the internal spaces are unobservable if their scales are of
order or less than the Fermi length $L_{\rm F} \sim
10^{-17}\mbox{cm} \sim 1\mbox{TeV}^{-1}$. Such small scales can be
achieved by two ways. First, the internal dimensions behave
dynamically toward the decrease of their size below $L_{\rm F}$.
Here, the internal spaces undergo dynamical evolution all the
time. This behavior is called dynamical compactification. Second,
the internal spaces can be stabilized near some fixed value, e.g.
$L_{\rm F}$. This behavior is called stable compactification.

For the first class of models (with dynamical compactification),
in Ref. \cite{Mohammedi} it was proposed an approach for the
reduction of multidimensional models with perfect fluid to an
effective four-dimensional ones which have the form of the
conventional cosmology. In the present paper we elaborate this
model and show that it provides very interesting gravitational
"constant" tuning effect. In spite of dynamical behavior of an
effective 4-D gravitational "constant" and a non-conventional
dynamics of an effective 4-D energy density, their product behaves
exactly as in FRW scenario. Precisely this product enters into
Friedmann equations. Thus, the external space has dynamical
evolution in accordance with the standard FRW cosmology. However,
it is well known that in KK approach dynamical internal spaces
lead to variation of the fundamental constants (see e.g.
\cite{Uzan} - \cite{GSZ} and references therein). As result, we
show that the fundamental constants in model \cite{Mohammedi}
undergo too large variations in comparison with experimental data.

Next, we consider models with stable compactification. It is worth
of noting that two particular classes of solutions with the stable
compactification of the internal spaces (for models with perfect
fluid) were already found in our paper \cite{CQG(1998)}. In the
present paper we prove a no-go theorem according to which the
models with perfect fluid do not admit the stable compactification
in the case of an arbitrary combination of equations of state in
the external and internal spaces\footnote{For a particular case,
this theorem was also confirmed in paper \cite{Gust,Bringmann}.} .
There are only two exceptional classes where the stable
compactification takes place and these classes exactly coincide
with those found in \cite{CQG(1998)}. Hence, these classes
entirely exhaust all possibilities for the stable
compactification. We construct a particular model which belongs to
these classes and has a Friedmann-like behavior for the external
space (our Universe) during the radiation and matter dominated
stages and late-time acceleration. However, the parameters of
model should be fine tuned to get the observable dark energy. The
reason of it is simple and rather common for such type of KK
models. From the stability condition (condition for the minimum of
an effective potential) follows that all parameters of the model
have the same order of magnitude as the curvature of the internal
space\footnote{As far as we know, the idea of TeV-scale extra
dimensions was first proposed  in \cite{Anton}.}: $R_1 \sim
L^{-2}_{F} \sim 10^{34}\mbox{cm}^{-2}$ which is much greater than
the observable value of the dark energy\footnote{The experimental
data give an estimate for the dark energy density: $\rho_{DE} \sim
10^{-123} \rho_{Pl} \Longrightarrow \Lambda_{DE} \sim
10^{-123}\Lambda_{Pl}$.} $\Lambda_{DE} \sim 10^{-123}\Lambda_{Pl}
\sim 10^{-57}\mbox{cm}^{-2}$. So, only with the help of extreme
fine tuning, the parameters can adjust to each other in such a way
to leave a fraction corresponding to $\Lambda_{DE}$.

To avoid the fine tuning problem, we generalize this model to the
case of orbifold internal space with branes in fixed points. In
the spirit of Universal Extra Dimension (UED) models \cite{UED},
the Standard Model fields are not localized on the brane but can
move in the bulk. We consider flat orbifolds. Therefore, the
curvature of the internal space is absent and the rest of
parameters can be of the order of $\Lambda_{DE}$ without fine
tuning. However, in this case the effective potential is too flat
(mass gravexcitons is very small) to provide necessary constancy
of the effective fundamental "constants".

The paper is structured as follows. In section 2, we explain the
general setup of our model and give a basic description of KK
models with multicomponent perfect fluid. Section 3 is devoted to
consideration of a model with dynamical compactification. In spite
of the Friedmann-like behavior, it is shown that effective
fundamental four-dimensional constants undergo too fast
variations. In section 4 we prove the no-go theorem on stable
compactification of internal spaces in model with perfect fluid
and find only two exceptional classes where such compactification
is possible. For these cases, in section 5, we construct a model
with conventional cosmology in the external space. Here, the
effective 4-D cosmological constant is positive. However,
parameters of the model should be fine tuned to get observable
dark energy density. In section 6, we generalize the latter model
to the case of stable orbifold. It gives us a possibility to avoid
the problem of fine tuning. However, the effective potential is
not curved enough to provide necessary constancy of the effective
fundamental "constants". The main results are summarized in the
concluding section 7.

\section{General setup}

\setcounter{equation}{0}

To start with, let us consider a cosmological model with
factorizable geometry,
\ba{1.1} g&=& \ov g^{\, (0)}(x)+\sum_{i=1}^n L_{Pl}^2 e^{2\beta
^i(\tau )}g^{(i)}(y) \\ \nn &\equiv & -e^{2\gamma (\tau )}d\tau
\otimes d\tau + L_{Pl}^2e^{2\beta ^0(\tau )}g^{(0)}(\vec{x}) +
\sum_{i=1}^n L_{Pl}^2 e^{2\beta ^i(\tau )}g^{(i)}(y) \, ,
\ea
which is defined on the manifold \rf{0.0} where, for generality,
$M_{D'}$ is also a direct product of $n$ compact $d_i$-dimensional
spaces: $M_{D'} = \prod\nolimits_{i=1}^n \mathcal{M}_i, \;
\sum\nolimits_{i=1}^n d_i = D'$. Metric $\ov g^{\, (0)}$ describes
the external (our) spacetime   $M_0 = \mathbb{R}\times
\mathcal{M}_0$ with dimension $D_0 = 1+d_0 = 4$. In order get
effective four-dimensional cosmology in the form of the Friedmann
equations, we assume that the factors $\mathcal{M}_i$ are Einstein
spaces: $R_{kl}[g^{(i)}] = \lambda^i g^{(i)}_{kl},\; i = 0,\ldots
,n , \; k,l = 1,\ldots ,d_i$ and $R[g^{(i)}]=\lambda^id_i \equiv
R_i$. In the case of constant curvature spaces parameters $\lambda
^i$ are normalized as $\lambda ^i=k_i(d_i-1)$ with $k_i=\pm 1,0$.
Quantities $a \equiv L_{Pl}e^{\beta^0}$ and $b_i \equiv
L_{Pl}e^{\beta^i} \, (i =1,\ldots ,n)$ describe scale factors of
the external and internal spaces, respectively.

With respect to the internal spaces, there are two possible
scenarios: either they are stably compactified at the present time
values $b_{(0)i} \equiv L_{Pl}e^{\beta^i_0} = \mbox{const}$, or
there is no such stabilization and $b_i$ remain dynamical
functions. In latter case, $\beta^i (\tau)$ define the averaged
dynamics of the volume of the internal spaces and $b_{(0)i}$ is
just instantaneous values of $b_i$ corresponding to the present
time $\tau_0$. For both of these cases, it make sense to consider
small inhomogeneous particle-like excitations/fluctuations
$\ov{\beta}^{\, i}(x)$ over such background (either over constant
$\beta^i_0$ or over dynamical $\beta^i (\tau)$), which describe
massive scalar particles (gravexcitons/radions) developing in the
external spacetime  \cite{PRD(1997),PRD(2000)}. The total volume
of the internal spaces at the present time is given by
\be{1.2} V_{D^{\prime }}
 \equiv V_I\times v_0 \equiv \prod_{i=1}^n\int\limits_{M_i}d^{d_i}y
\sqrt{|g^{(i)}|} \times \left( \prod_{i=1}^n e^{d_i\beta^i_0}
(L_{Pl})^{D'}\right) = V_I \times \prod_{i=1}^n b_{(0)i}^{d_i}.
\ee
The factor $V_I$ is dimensionless and defined by geometry and
topology of the internal spaces.

The action functional for considered multidimensional models reads
\be{1.4}
S=\frac 1{2\kappa_D ^2}\int\limits_Md^Dx\sqrt{|g|}\left\{
R[g]-2\Lambda_D \right\} + S_m\, , \ee
where $\kappa^2_D$ is $D$-dimensional fundamental gravitational
constant and $S_m$ is an action for bulk matter. In conventional
cosmology matter fields are taken into account in a
phenomenological way as a perfect fluid with equal pressure in all
three special directions. It provides homogeneous (if energy
density and pressure depends only on time) and isotropic picture
of the Universe. In multidimensional case we generalize this
approach to a $m-$component perfect fluid with energy-momentum
tensor
\be{1.5} T_N^M=\sum_{c=1}^m{T^{(c)}}^M_N ,
\ee
\be{1.6} {T^{(c)}}^M_N = {\rm diag\ } (\, -\rho ^{(c)}(\tau ),
\underbrace{P_0^{(c)}(\tau ),\ldots ,P_0^{(c)}(\tau )}_{\mbox{$d_0$ times}%
},\ldots , \underbrace{P_n^{(c)}(\tau ),\ldots ,P_n^{(c)}(\tau )}_{%
\mbox{$d_n$ times}}\, )\, . \ee
The conservation equations we impose on each component separately
\be{1.7} {T^{(c)}}^M_{N;M} = 0.
\ee
Denoting by an overdot differentiation with respect to time $\tau
$ in metric \rf{1.1}, these equations read for the tensors
(\ref{1.6})
\be{1.8}
\dot\rho ^{(c)}+\sum_{i=0}^nd_i\dot \beta ^i\left( \rho
^{(c)}+P_i^{(c)}\right) =0\, .
\ee
If the pressures and energy density are related via equations of
state
\be{1.9}
P_i^{(c)}=\left( \alpha _i^{(c)}-1\right) \rho ^{(c)},\ \ \ \ \
i=0,\ldots ,n,\quad c=1,\ldots ,m\, ,
\ee
then eq. \rf{1.8} have the simple integrals
\be{1.10}
\rho ^{(c)}(\tau )=A^{(c)}a^{-d_0 \alpha_0^{(c)}}\times
\prod_{i=1}^n b_i^{-d_i\alpha _i^{(c)}}\, ,
\ee
where $A^{(c)}$ are constants of integration.


\section{Dynamical compactification: fundamental "constant" variation problem}

\setcounter{equation}{0}

In this section we investigate a possibility for conventional
cosmology in the case of multidimensional models with dynamical
behavior of the internal spaces. More precisely, we consider the
case of dimensional stabilization when the scale factors of the
internal spaces $b_i$ decrease with time.

For simplicity, we consider the case of one internal space: $n=1,
\, b_1 \equiv b$. Perfect fluid is also taken in one component
form: $c=1 \Rightarrow \rho^{(1)} \equiv \rho, \, P_0^{(1)}\equiv
P_0, \, P_1^{(1)}\equiv P_1$. Then, the Einstein equations is
reduced to 3 differential equations:
\be{3.1}
\kappa^2_D \rho = \left\{3 H^2 + \frac{3k_0}{a^2}
-\Lambda_D\right\} + \frac12 d_1(d_1-1)\left(\frac{\dot
b}{b}\right)^2 + 3d_1 H\frac{\dot b}{b} +
\frac{d_1}{2}(d_1-1)\frac{k_1}{b^2}\, ,
\ee
\be{3.2}
\kappa^2_D P_0 = \left\{ -2 \frac{\ddot a}{a} -H^2
-\frac{k_0}{a^2} +\Lambda_D\right\} -d_1\frac{\ddot b}{b} -
\frac12 d_1(d_1-1)\left(\frac{\dot b}{b}\right)^2 -2d_1 H
\frac{\dot b}{b} - \frac{d_1}{2}(d_1-1)\frac{k_1}{b^2}
\ee
and
\ba{3.3} \kappa^2_D P_1 &=& -3 \frac{\ddot a}{a} -3 H^2 -
\frac{3k_0}{a^2} + \Lambda_D - (d_1-1)\frac{\ddot b}{b} -\frac12
(d_1-1)(d_1-2)\left(\frac{\dot b}{b}\right)^2 \\ \nn &-&3
(d_1-1)H\frac{\dot b}{b} - \frac12 (d_1-1)(d_1-2)
\frac{k_1}{b^2}\, ,
\ea
where $H \equiv \dot a/a$ is the Hubble parameter and dots denote
differentiation with respect to synchronous time $t$ (which
corresponds to the gauge $\gamma (\tau) = 0 \Rightarrow \tau
\equiv t$ in metric \rf{1.1}). This set of equations should be
supplemented with eq. \rf{1.8}, which results in solution
\rf{1.10} in the case of equations of state \rf{1.9}: $P_i=\left(
\alpha _i-1\right) \rho ,\; i=0,1$.

It worth of noting that eqs. \rf{3.1} and \rf{3.2} have the form
of the standard FRW equations if we leave in the rhs only terms in
curly brackets. So, we arrive to natural question. Could we
rewrite these equations in such a way that they reproduce
equations of the conventional cosmology, i.e. the Friedmann
equations? On the face of it, the answer is positive. The reason
for such answer is based on the following observation made in Ref.
\cite{Mohammedi}: if we suppose that the scale factor of the
internal space evolves according to the relation\footnote{In this
case the internal space decreases with expansion of the external
one. So, we obtain a model with dynamical compactification of the
extra dimensions. For such models the rate of compactification
plays very important role. This point will be discussed at the end
of this section.}
\be{3.4}
b = \frac{B}{a^q}\, , \quad B\equiv \mbox{const}
\ee
and parameter $q$ satisfies the condition
\be{3.5}
d_1q(d_1q-q -6) = 0
\ee
then, for the case of Ricci-flat internal space ($k_1 =0$), eqs.
\rf{3.1}--\rf{3.3} can be correspondingly rewritten as
\be{3.6}
\kappa_0^2 \rho_{(4)} = 3 H^2 + \frac{3k_0}{a^2} -\Lambda_D\, ,
\ee
\be{3.7}
\kappa_0^2 P_{(4)} = -2 \frac{\ddot a}{a} -H^2 - \frac{k_0}{a^2}
+\Lambda_D
\ee
and
\be{3.8}
\kappa_D^2 P_1 =  \left(d_1q -q -3\right)\frac{\ddot a}{a} -
\frac12 \left[6 +q(d_1-1)(d_1q-4)\right]H^2 -\frac{3k_0}{a^2} +
\Lambda_D\, .
\ee
Here,
\be{3.9}
\rho_{(4)} \equiv \rho \, V_{d_1}\, , \quad P_{(4)} \equiv
\left[P_0 - \frac{d_1q}{3}(\rho +P_1)\right]V_{d_1}
\ee
are identified with observable four-dimensional energy density and
pressure, respectively.

It can be easily seen, that eqs. \rf{3.6} and \rf{3.7} formally
reproduce the famous FRW equations (in the presence of the
cosmological constant). The third eq. \rf{3.8} can be used to find
the pressure along the extra dimensions in term of the scale
factor $a(t)$ of the observable Universe. However, there are two
main difference between this effective model and the standard FRW
universe. First of all, the effective four-dimensional
gravitational "constant" $\kappa_0^2$ is not a constant but a
dynamical function:
\be{3.10}
\kappa_0^2 = \kappa_D^2 / V_{d_1}\, .
\ee
In considered model, the volume of internal space reads
\be{3.11}
V_{d_1} = V_I b^{d_1} = V_I B^{d_1}/a^{d_1 q} \equiv   \ov V_I
a^{-d_1 q}\, .
\ee
It is worth of noting, that eq. \rf{3.5} is necessary condition
for a correct normalization of the effective four-dimensional
gravitational "constant". Only in this case we can reproduce eq.
\rf{3.10}. If parameter $q$ does not satisfy condition \rf{3.5},
it results in non-correct gauging of $\kappa_0$ (see corresponding
comments in paper \cite{remarks}). Eq. \rf{3.5} has two solutions:
$q=0$, which describes the static internal space, and
\be{3.12}
q= \frac{6}{d_1-1}\, , \quad d_1 \ne 1\, ,
\ee
which corresponds to the dynamical compactification. In this
section we shall concentrate on the latter case mentioning briefly
the static case at the end of the section.

Second very important difference between our effective model and
conventional cosmology consists in the equation of conservation of
energy. For our model, eq. \rf{1.8} with the help of eqs. \rf{3.9}
can be written in the following form:
\be{3.13}
\frac{d}{dt} \left(a^3 \rho_{(4)}\right) + P_{(4)}
\frac{d}{dt}\left(a^3\right) = \left(a^3 \rho_{(4)}\right)
\frac{1}{V_{d_1}}\frac{d}{dt}\left(V_{d_1}\right) = - \left(a^3
\rho_{(4)}\right)d_1 q \frac{\dot a}{a}\, ,
\ee
which, obviously, differs from the standard conservation equation
in the FRW Universe by non-zero rhs.

Similar to the FRW Universe, it can be easily shown that eq.
\rf{3.7} comes out from eq. \rf{3.6} and \rf{3.13}. To prove it,
it is necessary to keep in mind that $\kappa_0$ is dynamical
function \rf{3.10}. So, there is no need to solve eq. \rf{3.7}.

Let us suppose now that the energy density $\rho_{(4)}$ and
pressure $P_{(4)}$ are connected with each other via equation of
state:
\be{3.14}
P_{(4)} = (\alpha - 1)\rho_{(4)}\, .
\ee
Then, eq. \rf{3.13} has the following solution:
\be{3.15}
\rho_{(4)} = \rho_{0}\left(\frac{a_0}{a}\right)^{3\alpha + d_1q}\,
,
\ee
where $\rho_0$ is the energy density at the moment when scale
factor is $a_0$. This behavior of the energy density differs from
the standard one by additional degree $d_1q$, which varies from 6
to 12 if dimension $d_1$ varies from $d_1 \gg 1$ to $d_1=2$,
respectively. However, it is very important to note, that
combination
\be{3.16}
\kappa_0^2 \rho_{(4)} \sim a^{-3\alpha}
\ee
as for the standard cosmology. It follows from the fact that
dynamical behavior of $\kappa_0^2 \sim a^{d_1q}$ exactly
compensates additional degree $a^{-d_1q}$ in expression for
$\rho_{(4)}$. As a result, dynamical evolution of the Universe in
our model exactly coincides with evolution of the standard FRW
Universe. For example, if we take a flat Universe ($k_0 = 0$)
without cosmological constant $\Lambda_D =0$, then the solution of
eq. \rf{3.6} has the standard for FRW Universe form:
\be{3.17}
a = \left(\frac{3\alpha }{2} a_{\ast}t\right)^{2/(3\alpha )}\, ,
\ee
where $a_{\ast} \equiv \kappa_D^2 \rho_0 a_{0}^{3\alpha +
d_1q}/(3\ov V_I)$. Hence, the combination $\kappa_0^2 \rho_{(4)}
\sim t^{-2}$ has conventional cosmological behavior.

Thus, we recover the standard behavior of the Universe in our
multidimensional model due to specific dynamic of the effective
four-dimensional gravitational constant $\kappa_0^2 \equiv 8\pi
G_4 \Longrightarrow G_4 \sim V_{d_1}^{-1} \sim a^{d_1q}$ (see eq.
\rf{3.10}). Contrary to the Dirac hypothesis on $G_4 \sim t^{-1}$
and negative $\dot G_4/G_4 <0$ \cite{Dirac}, in our model $G_4$
increases with time (if the scale factor $a$ is a growing function
of time) and $\dot G_4/G_4$ is positive. For example, in the case
of solution \rf{3.17} we get
\be{3.18} G_4 \sim t^{2d_1q/(3\alpha)}\quad \Longrightarrow \quad
\frac{\dot G_4}{G_4}= \frac{2d_1q}{3\alpha}\frac{1}{t}\, > 0, \;
\forall \, \alpha
>0\, .
\ee
It gives for the age of the Universe $t_u \sim 14$Gyr
\be{3.19}
\frac{\dot G_4}{G_4} \sim 10^{-10}{\mbox{yr}}^{-1}\, .
\ee

There are a number of estimates for possible variations of $G_4$
(see e.g. \cite{Uzan}-\cite{DIKM} and references therein).
Comparison with these estimates shows that our value \rf{3.19} is
too large\footnote{Similar conclusion on too large variations of
the 4-D gravitational constant in multidimensional models with
ideal perfect fluid and dynamical internal spaces was obtained in
paper \cite{CV}.}. $|\dot G_4/G_4|$ should be smaller by at least
two orders of magnitude. Negative sign of $\dot G_4/G_4$ is also
preferable although positive values are not excluded
\cite{Kubyshin}. In spite of rather large value \rf{3.19},
it is worth of noting that some of estimates are not applicable
for our model. The reason of it is that the Universe in our model
has the same rate of evolution as the conventional Universe. Thus,
the time-depending $G_4$ in our case has no direct effects on the
CMB angular power spectrum as well as on the nucleosynthesis.

{}From other hand, dynamical behavior of the internal spaces can
result in variation others fundamental constants. For example, the
inclusion in the model of the electromagnetic fields leads to the
variation of the fine structure constant: $\alpha_4 = \alpha_D
/V_{d_1}$ \cite{Kubyshin,GSZ}. Thus, variations of $G_4$ and
$\alpha_4$ are connected with each other as follows:
\be{3.20}
\frac{\dot G_4}{G_4} = \frac{\dot \alpha_4}{\alpha_4}\, .
\ee
The observation of quasar absorbtion lines shows that $\dot
\alpha_4 /\alpha_4 \sim 10^{-15}{\mbox{yr}}^{-1}$ \cite{Webb}.
Thus, provided that these observational determination of $\dot
\alpha /\alpha$ is correct, we should also have this estimate for
$\dot G_4/G_4$. Obviously, this is much less than \rf{3.19}. The
relatively large value of $\dot G_4/G_4$ in our model originates
from a high rate of evolution of the internal space: $V_{d_1} \sim
a^{-d_{1}q} \sim t^{-2d_1q/(3\alpha )} = t^{-4d_1/[\alpha
(d_1-1)]}$. For radiation ($\alpha = 4/3$) and dust ($\alpha =
1$), we have correspondingly  $3< 2d_1q/(3\alpha ) \leq 6$ and $4<
2d_1q/(3\alpha ) \leq 8$.

The evolution of the internal space can be slow down if parameter
$d_1q \rightarrow 0$. However, its value is defined by eq.
\rf{3.12} and varies in limits from 6 to 12. We can resolve this
problem taking the second solution of eq. \rf{3.5}: $q=0$, which
corresponds to the static internal space: $b \equiv \mbox{const}$.
Unfortunately, it is difficult to justify this solution for
arbitrary $\rho , P_0$ and $P_1$. Other words, solution $b =
\mbox{const}$ is, in general, not stable (for details, see the
next section). Nevertheless, for some particular cases, stable
solutions exist and we discuss them in the next section.


\section{Stable compactification: no-go theorem}

\setcounter{equation}{0}

To investigate the problem of the stable compactification, it is
helpful to use the equivalence between the Einstein eqs. \rf{3.1}
- \rf{3.3} and the Euler-Lagrange equations for Lagrangian
obtained by dimension reduction of the action \rf{1.4} with
\be{4.1}
S_m = - \int\limits_Md^Dx\sqrt{|g|} \rho\, ,
\ee
where $\rho $ is given by eq. \rf{1.10} (see \cite{CQG(1996)} for
details). This equivalence takes place for homogeneous model
\rf{1.1}.  However, we can generalize it to the inhomogeneous case
allowing inhomogeneous fluctuations $\ov
\beta^i (x)$ over stably compactified background $\beta_0^i =
\mbox{const}$:
\be{4.2}
\ov \beta^i (x) = \beta^i (x) - \beta_0^i\, , \quad i=1,\ldots
,n\; ,
\ee
where coordinates $x$ are defined on the manifold $M_0$. In this
section, $b_{(0)i} = L_{Pl} \exp{\beta_0^i}$ are treated as the
scale factors of the internal spaces stabilized at the present
time.

Then, after conformal transformation of the external spacetime
metric from the Brans-Dicke to the Einstein frame\footnote{Most
easily the analysis of the internal space stabilization can be
done in the Einstein frame. For this purpose we perform this
conformal transformation. Evidently, if the stabilization takes
place in the Einstein frame (i.e. $\ov \beta ^i = 0$), it also
occurs in the BD frame because both of these frames in this case
coincide with each other.}:
\be{4.3}
\ov g^{(0)}_{\mu \nu}(x) = \Omega^2 \tilde g^{(0)}_{\mu \nu}(x) :=
{\left( \prod_{i=1}^ne^{d_i\ov \beta ^i}\right) }
^{\frac{-2}{D_0-2}} \tilde g_{\mu \nu }^{(0)}\, ,
\ee
the dimensional reduction of action \rf{1.4} with the matter term
\rf{4.1} results in the following four-dimensional effective
theory (see for details, \cite{CQG(1998),PRD(1997),PRD(2000)}):
\be{4.4}
S=\frac 1{2\kappa _0^2}\int\limits_{ M_0}d^{D_0}x\sqrt{%
|\tilde g^{(0)}|}\left\{ \tilde R\left[ \tilde g^{(0)}\right]
-\bar G_{ij}\tilde g^{(0)\mu \nu }\partial _\mu \ov \beta
^i\,\partial _\nu \ov \beta ^j-2U_{eff}\right\}\, ,
\ee
where $\kappa_0^2 = \kappa_D^2 / V_{D'}$ is effective
four-dimensional gravitational constant with $V_{D'}$ from eq.
\rf{1.2} (with $b_{(0)i} = L_{Pl} \exp{\beta_0^i} = \mbox{const}$)
and corresponds to the present day value (i.e. to the Newton
gravitational constant). In eq. \rf{4.4}, $\bar G_{ij} =
d_i\delta_{ij} + [1/(D_0-2)]d_id_j$ and
\be{4.5}
U_{eff}={\left( \prod_{i=1}^ne^{d_i\ov \beta ^i}\right) }^{-\frac
2{D_0-2}}\left[ -\frac 12\sum_{i=1}^n\tilde R_ie^{-2\ov \beta
^i}+\Lambda_D +\kappa ^2_D\sum_{c=1}^m\rho ^{(c)}\right]\, ,
\ee
is the effective potential where $\tilde R_i := R_i
L_{Pl}^{-2}e^{-2 \beta_0^i}$ and $\rho^{(c)}$ is defined by eq.
\rf{1.10}. If we suppose that the external spacetime  metric in
the Einstein frame has also the FRW form:
\be{4.6}
\tilde g^{(0)}=\Omega ^{-2}\ov g^{(0)}=\tilde g_{\mu \nu
}^{(0)}dx^\mu \otimes dx^\nu :=-e^{2\tilde \gamma }d\tilde \tau
\otimes d\tilde \tau +L_{Pl}^2 e^{2\tilde \beta ^0(x)}g^{(0)}\, ,
\ee
which results in the following connection between the external
scale factors in the Brans-Dicke frame $a\equiv L_{Pl}e^{\beta
^0}$ and in the Einstein frame $\tilde a \equiv L_{Pl}e^{\tilde
\beta ^0}$:
\be{4.7}
a={\left( \prod_{i=1}^ne^{d_i\ov \beta ^i}\right) }^{-\frac
1{D_0-2}}\tilde a\, ,
\ee
then, expression \rf{1.10} for $\rho^{(c)}$ can be rewritten in
the form:
\be{4.8}
\kappa^2_D \rho ^{(c)}=
\kappa^2_0 \rho^{(c)}_{(4)}\prod_{i=1}^n e^{-\xi_i^{(c)}\ov
\beta^i}\, ,
\ee
where
\be{4.9}
\rho^{(c)}_{(4)} = \tilde A^{(c)} \tilde a^{-d_0\alpha_0^{(c)}}\,
, \quad \tilde A^{(c)} = A^{(c)} V_I \prod_{i=1}^n
b_{(0)i}^{d_i(1-\alpha_i^{(c)})}
\ee
and
\be{4.10}
\xi_i^{(c)} = d_i\left( \alpha _i^{(c)}-\frac{\alpha _0^{(c)}d_0}{%
d_0-1}\right)\, .
\ee
It can be easily verified that $\tilde A^{(c)}$ has dimension
$\mbox{cm}^{d_0\alpha_0^{(c)}-D_0}$.

Thus, the problem of stabilization of the extra dimensions is
reduced now to search of minima of the effective potential
$U_{eff}$ with respect to the fluctuations $\ov \beta^i$:
\be{4.11}
\left.\frac{\partial U_{eff}}{\partial \ov \beta^k}\right|_{\ov
\beta =0} =0 \Longrightarrow \tilde R_k = -\frac{d_k}{D_0-2}\left[
\sum_{i=1}^n \tilde R_i -2\Lambda_D \right] + \kappa^2_0
\sum_{c=1}^m \rho^{(c)}_{(4)}\left(\xi^{(c)}_k +
\frac{2d_k}{D_0-2}\right) \, , \; k = 1,\ldots ,n\, .
\ee
The left-hand side of this equation is a constant but the
right-hand side is a dynamical function because of dynamical
behavior of the effective four-dimensional energy density
$\rho^{(c)}_{(4)}$ . Thus, we arrived to the following \\

{\bf No-go theorem:}\\


{\it Multidimensional cosmological Kaluza-Klein models with the
perfect fluid as a matter source do not admit stable
compactification of the internal spaces with exception of two
special cases:}

\ba{4.12}
\mbox{I.} \quad \alpha^{(c)}_0 &=& 0\, , \quad \forall \;
\alpha_i^{(c)} ,
\quad i=1,\ldots ,n,\quad c=1,\ldots ,m\, .\\
&\vphantom{\int}& \nn \\
\mbox{II.}\quad
\xi_i^{(c)}&=&-\frac{2d_i}{d_0-1} \vphantom{\int} \Longrightarrow
\left\{\begin{array}{rcl} \alpha _0^{(c)}&=&\frac
2{d_0}+ \frac{d_0-1}{d_0}\alpha ^{(c)}\, ,\\
&\vphantom{\int}&\\
\alpha _i^{(c)}&=&\alpha ^{(c)},\quad i=1,\ldots ,n,\quad
c=1,\ldots ,m\, .\\\end{array}\right.
\ea

\vspace{0.3cm}

\noindent First case corresponds to vacuum in the external space
$\rho^{(c)}_{(4)} = \tilde A^{(c)}=\mbox{const}$ and arbitrary
equations of state in the internal spaces. Some bulk matter can
mimic such behavior, e.g. vacuum fluctuations of quantum fields
(Casimir effect) \cite{PRD(1997),GKZ}, monopole form fields
\cite{PRD(1997),GMZ} and gas of branes \cite{Kaya}.

In the second case\footnote{In paper \cite{CQG(1998)}, it was
found that condition (4.13) results in effective potentials
\rf{4.5} with separating scale factor contributions from internal
and external factor spaces. It was stressed that this separation
crucially simplifies the analysis of the internal space stable
compactification. We can rewrite (4.13) in the form
$2-d_0\alpha_0^{(c)}+(d_0-1)\alpha_i^{(c)} =0$. In the case
$d_0=3, \, i=c=1$, the same expression was found in \cite{Gust}
where it was shown that it provides the static flat internal space
in the model with $\Lambda_D=0$. It can be easily shown that such
static solution is unstable.}, the energy density in the external
space is not a constant but a dynamical function with the
following behavior\footnote{The corresponding equation of state
is: $P_{(4)}^{(c)} = (1/3)(2\alpha^{(c)}-1)\rho _{(4)}^{(c)}$,
where we put $d_0=3$.}:
\be{4.15}
\rho _{(4)}^{(c)}(\tilde a)=\tilde A^{(c)}\frac 1{\tilde
a^{2+(d_0-1)\alpha ^{(c)}}} =\left. \tilde A^{(c)}\frac 1{\tilde
a^{2(1 +\alpha ^{(c)})}}\right|_{d_0=3}\, .
\ee
For example, in three-dimensional external space, such perfect
fluid has the form of a gas of cosmic strings for $\alpha
^{(c)}=0$, dust for $\alpha ^{(c)}=1/2$ and radiation for $\alpha
^{(c)}=1$.


\section{Late time acceleration of the Universe: fine tuning problem}

\setcounter{equation}{0}

Let us try now to built more o less viable model with stabilized
internal spaces. In our paper \cite{CQG(1998)} were have already
investigated Friedmann-like multidimensional cosmological models
with the perfect fluid of \rf{4.15}. It was shown that one of the
necessary conditions for the stable compactification is the
negativeness  of an effective four-dimensional cosmological
constant. However, a negative cosmological constant leads to a
deceleration of the Universe instead of an accelerated expansion,
as recent observational data indicate \cite{SnIa}. In
\cite{GMZ(1)} we already indicated that the effective cosmological
constant can be shifted from negative values to positive ones by
including into the nonlinear model matter fields, e.g. additional
form fields. This hypothesis was confirmed in Ref. \cite{GMZ}. In
the present paper, we try to achieve the similar effect combining
together cases I. and II. To be more precise, additionally to the
the perfect fluid of the type II, we endow our model with the
monopole form fields \cite{PRD(1997),GMZ}:
\be{4.14}
S_m = -\frac12 \int_M d^Dx \sqrt{|g|}\sum_{i=1}^n \frac{1}{d_i!}
\left(F^{(i)} \right)^2 = - \int_M d^Dx \sqrt{|g|}\sum_{i=1}^n
\frac{f_i^2}{b_i^{2d_i}}\, ,
\ee
where $f_i \equiv \mbox{const}$ are arbitrary constants of
integration (free parameters of the model) and for real form
fields $f^2_i > 0$. Comparison of this expression with eqs.
\rf{4.1} and \rf{1.10} shows that such monopole form fields are
equivalent to n-component perfect fluid with $\alpha_0^{(c)} =0 ,
\; \alpha^{(c)}_i = 2\delta^c_i ,\quad c,i = 1,\ldots ,n$, i.e.
belong to the case I.

Without loss of prediction power of our conclusions, we can
perform our analysis in the case of one internal space $n=1$.
Then, the effective potential for such combined model\footnote{We
consider the case where the Casimir energy density is negligible
in comparison with the energy density of the form fields.}
undergos the following separation:
\be{4.16} U_{eff}=\stackunder{U_{int}(\ov \beta{\, ^1})}{\underbrace{{\left(
e^{d_1\ov \beta{\, ^1}}\right) }^{-\frac 2{D_0-2}}\left[ -\frac
12\tilde R_1e^{-2\ov \beta{\, ^1}}+\Lambda_D + \tilde
f^{2}_1e^{-2d_1\ov \beta^{\, 1}} \right] }}\
+\stackunder{U_{ext}(\tilde a)}{ \underbrace{\kappa_0
^2\sum_{c=1}^m\rho _{(4)}^{(c)}(\tilde a)},}
\ee
where $\rho _{(4)}^{(c)}(\tilde a)$ is defined by eq. \rf{4.15}
and $\tilde f^2_1 \equiv \kappa^2_D f^2_1 / b_{(0)1}^{2d_1}$. This
separation is the result of cancellation of the prefactor
$\exp{(-2d_1\ov \beta ^1/(D_0-2))}$ in $U_{eff}$ \rf{4.5} with
corresponding prefactor in effective 4D energy density form the
case II: $\kappa^2_{D}\rho^{c} = \kappa^2_0 \rho^{(c)}_{(4)}
\exp{(-\xi_1^{(c)}\ov \beta ^1)} =  \kappa^2_0 \rho^{(c)}_{(4)}
\exp{(2d_1\ov \beta ^1/(D_0-2))}$. We will show below, that such
separation on the one hand provides a stable compactification of
the internal factor space due to a minimum of the first term
$U_{int}=U_{int}(\ov
\beta{\, ^1})$ as well as a dynamical behavior of the external
factor space due to $U_{ext}=U_{ext}(\tilde a)$.

First, we investigate the problem of stable compactification of
the internal space. It is clear that such stabilization for our
model takes place if potential $U_{int}$ has a minimum with
respect to fluctuation field $\ov \beta^{\, 1}$:
\be{4.17}
\left.\frac{\partial U_{int}}{\partial \ov \beta^{\, 1
}}\right|_{\ov \beta^{\, 1} =0} =0 \Longrightarrow
\frac{D-2}{2d_1}\, \tilde R_1 = \Lambda_D +d_0\tilde f^{2}_1\, .
\ee
The value of this potential at the minimum plays the role of
effective four-dimensional cosmological constant:
\be{4.18}
\Lambda_{eff} := \left. U_{int}\right|_{\ov \beta^{\, 1}=0} =
-\frac12 \tilde R_1 +\Lambda_D + \tilde f^2_1\, .
\ee
With the help of the extremum condition \rf{4.17}, $\Lambda_{eff}$
can be written in the form
\ba{4.19} \Lambda_{eff} &=& \frac{D_0-2}{2d_1}\, \tilde R_1 - (D_0-2)
\tilde f^2_1  \\ &=& \frac{D_0-2}{D-2} \Lambda_D - \left(
\frac{d_0 d_1}{D-2} - 1 \right)\tilde f^2_1 \label{4.20} \\ &=&
\frac{d_0-1}{d_0}\Lambda_D - \frac12  \left(1 -
\frac{D-2}{d_0d_1}\right)\tilde R_1 \label{4.21}\, .
\ea
Second derivative of $U_{int}$ in the extremum position reads
\ba{4.22}
\left.\frac{\partial^2 U_{int}}{\partial {\ov \beta^{\, 1
}}^2}\right|_{\ov \beta^{\, 1} =0} &=&
-2\left(\frac{D-2}{D_0-2}\right)^2\tilde R_1 +
\left(\frac{2d_1}{D_0-2}\right)^2\Lambda_D +
\left(\frac{2d_0d_1}{D_0-2}\right)^2\tilde f^2_1 \\
&=& \frac{4}{D_0-2}\left[-\frac12 (D-2)\tilde R_1 +
4d_0d_1^2\tilde f^2_1\right]\label{4.23}\\
&=& \frac{4d_1}{(D_0-2)^2}\left[-(D_0-2)\Lambda_D + (D-2)d_0
\left(\frac{d_0d_1}{D-2}-1\right)\tilde f^2_1 \right]\label{4.24}\\
&=& \frac{4}{(D_0-2)^2}\left[-d_1^2 (d_0-1)\Lambda_D +
\frac12(D-2)^2\left(\frac{d_0d_1}{D-2}-1\right)\tilde
R_1\right]\label{4.25}\, .
\ea
For stable compactification, this extremum should be a minimum.
Then, small fluctuations above it describe minimal scalar field
(gravitational excitons \cite{PRD(1997)}) propagated in the
external space with the mass squared
\be{4.26}
m^2_{exci} \equiv \frac{D_0-2}{d_1(D-2)}\left.\frac{\partial^2
U_{int}}{\partial {\ov \beta^{\, 1 }}^2}\right|_{\ov \beta^{\, 1}
=0} >0\, .
\ee
Additionally, the effective four-dimensional cosmological constant
should be positive $\Lambda_{eff} > 0$. This is the necessary
condition for the late time acceleration of the Universe in
considered model. Both of these conditions (the positiveness of
$\Lambda_{eff}$ and $m^2_{exci}$) lead to the following
inequalities\footnote{The inequalities in the lhs result from the
condition $\Lambda_{eff} > 0$ applied to eqs. \rf{4.19}
-\rf{4.21}, whereas the inequalities in the rhs follow from the
minimum condition $m^2_{exci} >0$
applied to eqs. \rf{4.23} - \rf{4.25}. }
\ba{4.27}
d_1 \tilde f^2_1 < &\frac12 \tilde R_1& <
\frac{4d_0d_1}{D-2}\times d_1\tilde f^2_1\\
\frac{D-2}{D_0-2}\left(\frac{d_0d_1}{D-2}-1\right)\tilde f^2_1 <
&\Lambda_D& < d_0\times
\frac{D-2}{D_0-2}\left(\frac{d_0d_1}{D-2}-1\right)\tilde f^2_1
\label{4.28}\\
\frac12
\frac{D-2}{d_1(d_0-1)}\left(\frac{d_0d_1}{D-2}-1\right)\tilde R_1
< &\Lambda_D& < \frac{D-2}{d_1}\times \frac12
\frac{D-2}{d_1(d_0-1)}\left(\frac{d_0d_1}{D-2}-1\right)\tilde R_1
\label{4.29}\, .
\ea
These inequalities clearly show that the positive minimum takes
place only if signs of  $\tilde R_1$ and $\Lambda_D $ are
positive: $\tilde R_1, \Lambda_D > 0$ (for real form fields with
$\tilde f^2_1 >0$). Thus, for non-positive $\tilde R_1$ and
$\Lambda_D$ the effective cosmological constant is also
non-positive\footnote{It is worth of noting that expression
$[(D-2)/(2d_1)]\, \tilde R_1 - \Lambda_D$ should be non-negative
for any signs of $\tilde R_1$ and $\Lambda_D $, as it follows from
eq. \rf{4.17}.}. For example, in the case of Ricci-flat ($\tilde
R_1 = 0$) internal spaces (e.g. $d_1=1$) the effective
four-dimensional cosmological constant can admit only negative
values.

According to the present day observations, our Universe undergoes
the late time accelerating expansion due to a dark energy
\cite{SnIa}. The origin of the dark energy is the great challenge
of the modern theoretical physics and cosmology. The cosmological
constant is one of the most probable candidate for it. According
to the observations, its value should be $\Lambda_{DE} \sim
10^{-123}\Lambda_{Pl} \sim 10^{-57}\mbox{cm}^{-2}$. Let us
estimate now a possibility for our effective cosmological constant
to admit this quantity: $\Lambda_{eff} \sim \Lambda_{DE} \sim
10^{-57}\mbox{cm}^{-2}$.

It is well known that in KK models
the size of extra dimensions at present time should be of order or
less than $b_{(0)1} \sim 10^{-17}\mbox{cm} \sim 1\mbox{TeV}^{-1}$.
In this case $\tilde R_1 \sim b_{(0)1}^{-2} \sim
10^{34}\mbox{cm}^{-2}$. From other hand, inequalities \rf{4.27} -
\rf{4.29} show that $\tilde R_1, \Lambda_D$ and $\tilde f^2_1$ are
of the same order of magnitude, i.e. $\tilde R_1 \sim \Lambda_D
\sim \tilde f^2_1 \sim 10^{34}\mbox{cm}^{-2}$, and have the same
sign. Thus, these parameters should be extremely fine tuned (in
eq. \rf{4.18}) to compensate each other in such a way that to
leave only $10^{-57}\mbox{cm}^{-2}$. We see two possibilities to
avoid this problem. First, the inclusion of different form
fields/fluxes may result in a big number of minima (landscape)
\cite{landscape} with sufficient large probability to find oneself
in a dark energy minimum. This problem we shall investigate in our
forthcoming paper. Second, we can avoid the restriction $\tilde
R_1 \sim b_{(0)1}^{-2} \sim 10^{34}\mbox{cm}^{-2}$ if the internal
space is Ricci-flat: $\tilde R_1 = 0$. For example,
$\mathcal{M}_1$ can be an orbifold with branes in fixed points
\cite{UED}. This model we investigate in the next section.

For masses of gravexcitons we obtain (if $b_{(0)1} \sim
10^{-17}\mbox{cm} \sim 1\mbox{TeV}^{-1}$)
\be{4.30}
m_{exci} \sim 1\mbox{TeV}\, .
\ee
Masses of such heavy gravexcitons are very close to values which
do not contradict to the observable data \cite{GSZ,iwara}. They
decay (e.g. due to reaction $\psi \rightarrow 2\gamma$) long
before the present time and the effective potential contributes in
the form of the effective cosmological constant \rf{4.18}.

Let us now turn to the dynamical behavior of the external factor
space. We consider zero order approximation, when all excitations
are freezed out (or heavy enough to decay before the present
time). Because our Universe (external space) is homogeneous and
isotropic, functions $\tilde \gamma $ and $\tilde \beta^0$ depends
only on time: $\tilde \gamma =\tilde \gamma (\tilde \tau )$ and
$\tilde
\beta ^0=\tilde
\beta^0 (\tilde \tau )$. Then, the action functional \rf{4.4} (for
combined model \rf{4.16}) after dimensional reduction reads:
\begin{eqnarray}
\label{4.31}
S & = & \frac 1{2\kappa _0^2}\int\limits_{\bar M_0}d^{D_0}x
\sqrt{|\tilde g^{(0)}|}\left\{ \tilde R\left[ \tilde
g^{(0)}\right]
-2U_{eff}\right\} =\nonumber \\
& = & \frac{V_0}{2\kappa _0^2}%
\int d\tilde \tau \left\{ e^{\tilde \gamma +d_0\tilde \beta^0
}e^{-2\tilde \beta^0 }R[g^{(0)}]+d_0(1-d_0)e^{-\tilde \gamma
+d_0\tilde \beta^0 }\left( \frac{d\tilde \beta^0 }{d\tilde \tau
}\right) ^2
\right. \nonumber \\
& & \left. -2e^{\tilde \gamma +d_0\tilde \beta^0 }\left( \Lambda
_{eff}+ \kappa_0 ^2\sum_{c=1}^m\rho _{(4)}^{(c)}\right) \right\} +
\frac{V_0}{2\kappa _0^2}d_0\int d\tilde \tau \frac d{d\tilde \tau
}\left(
e^{-\tilde \gamma +d_0\tilde \beta^0 }\frac{d\tilde \beta^0 }{d\tilde \tau }%
\right) ,
\end{eqnarray}
where $V_0 := \int_{\mathcal{M}_0}d^{d_0}x \sqrt{|g^{(0)}|}$,
$\rho _{(4)}^{(c)}$ is defined by eq. \rf{4.15} and usually
$R[g^{(0)}]=kd_0(d_0-1),\; k=\pm 1,0$. The constraint equation
$\partial L/\partial \tilde \gamma =0$ in the synchronous time
gauge $\tilde \gamma =0$ yields
\begin{equation}
\label{4.32}\left( \frac 1{\tilde a}\frac{d\tilde a}{d\tilde
t}\right) ^2=-\frac k{\tilde a^2}+\frac 2{d_0(d_0-1)}\left(
\Lambda _{eff}+ \kappa_0 ^2\sum_{c=1}^m\rho _{(4)}^{(c)}(\tilde
a)\right) ,
\end{equation}
which results in
\begin{eqnarray}
\label{4.33}
\tilde t+const & = & \int \frac{d\tilde a}{\left[
-k+\frac{2\Lambda _{eff}}{d_0(d_0-1)}\tilde a^2+ \frac{2\kappa_0
^2}{d_0(d_0-1)}\sum_{c=1}^m\frac{\tilde A^{(c)}}{\tilde
a^{(d_0-1)\alpha ^{(c)}}}\right] ^{1/2}}\ , \nonumber \\
& = & \int \frac{d\tilde a}{\left[ -k +\frac{\Lambda
_{eff}}3\tilde a^2+\frac{\kappa_0^2}3\sum_{c=1}^m \frac{\tilde
A^{(c)}}{\tilde a^{2\alpha ^{(c)}}}\right] ^{1/2}},
\end{eqnarray}
where in the last line we put $d_0=3$.

Thus in zero order approximation we arrived at a Friedmann model
in the presence of positive cosmological constant $\Lambda
_{eff}>0$ and a multicomponent perfect fluid. It is assumed that
$\Lambda _{eff}$ defines dark energy observed now.  The perfect
fluid has the form of a
dust for $\alpha ^{(c)}=1/2$ and radiation for $\alpha ^{(c)}=1$.
As for ordinary matter $\alpha ^{(c)} >0$ , the cosmological
constant plays a role only for large $\tilde a$ and because of the
positive sign of $\Lambda _{eff}$ the Universe undergos the late
time acceleration. The inclusion of gravexcitons into
consideration (the first order approximation) does not change this
picture because gravexcitons with masses \rf{4.30} decay into
usual matter before primordial nucleosynthesis \cite{GSZ}.

Within the bounds of this scenario, there is also a possibility
for a primordial inflation. For this purpose we can consider one
component perfect fluid with $\alpha^{(1)} <0$, e.g. $\alpha^{(1)}
= -1/2 \Rightarrow \alpha_0^{(1)}= 1/3$ which describes a
frustrated network of domain walls in the external space. It is
well known that such perfect fluid results in acceleration of the
Universe. For example, in the case $\alpha^{(1)} = -1/2$ the flat
Universe ($k=0$) undergoes the power law inflation at early times:
$\tilde a \sim \tilde t^2$. If domain walls decay into ordinary
matter, then the described above Friedmann-like behavior follows
the inflation.


\section{Stabilization of orbifolds}

\setcounter{equation}{0}

In this section, to avoid the fine tuning problem, we consider
Ricci-flat internal spaces. In this case scalar curvatures of the
internal spaces are absent and there is no need for extreme fine
tuning of the parameters to get the observable dark energy.

Among such models, Universal extra dimension models (UED) are of
special interest \cite{UED}. Here, the internal spaces are
orbifolds\footnote{For example, $S^1/Z_2$ and $T^2/Z_2$ which
represent circle and square folded onto themselves due to $Z_2$
symmetry.} with branes in fixed points. The compactification of
the extra dimensions on orbifolds has a number of very interesting
and useful properties, e.g. breaking (super)symmetry and obtaining
chiral fermions in four dimensions (see e.g. paper by H.-C. Cheng
at al in \cite{UED}). The latter property gives a possibility to
avoid famous no-go theorem of KK models (see e.g. \cite{no-go}).

In UED models, the Standard Model fields are not localized on the
brane but can move in the bulk. Branes in fixed points contribute
in action functional \rf{1.4} in the form:
\be{5.1}
\sum_{\phantom{x}^{fixed}_{points}}\left. \int d^4x \sqrt{\ov
g^{(0)}(x)} L_b \right|_{\phantom{x}^{fixed}_{point}}\, ,
\ee
where $\ov g(x)$ is induced metric (which for our geometry
\rf{1.1} coincides with the metric of the external spacetime  in
the Brans-Dicke frame) and $L_b$ is the matter Lagrangian on the
brane. In what follows, we consider the case where branes are only
characterized by their tensions $L_{b(k)} = -\tau_{(k)}\, ,\,
k=1,2,\ldots ,m$ and $m$ is the number of branes.

After transformation to the Einstein frame \rf{4.3}, the action
\rf{5.1} reads
\be{5.2}
\frac{1}{2\kappa^2_0} \int d^4x \sqrt{\tilde g ^{(0)}(x)}
\left(\prod_{i=1}^n e^{d_i \ov
\beta^{\, i}}\right)^{-\frac{2}{D_0-2}}\left[-2\kappa^2_0 \sum_{k=1}^m \tau_{(k)}
\prod_{i=1}^ne^{-d_i\ov \beta^{\, i}}\right]\, .
\ee
The comparison of this expression with eq. \rf{4.5} shows that
branes contribute in the effective potential in the form of one
component perfect fluid ($c=1$) with equations of state:
$\alpha^{(1)}_0 =0,\, \alpha^{(1)}_i = 1, \, i=1,\ldots ,n$, i.e.
from the case I of the no-go theorem. It means that they
contribute only to the $U_{int}$:
\be{5.3}
U_{int} = {\left( e^{d_1\ov \beta{\, ^1}}\right) }^{-\frac
2{D_0-2}}\left[ \Lambda_D + \tilde f^{2}_1e^{-2d_1\ov \beta^{\,
1}} - \lambda e^{-d_1\ov \beta^{\, 1}}\right]\, ,
\ee
where we consider the case of one internal space $i=1$ and
introduce notation $\lambda \equiv - \kappa^2_0 \sum_{k=1}^m
\tau_{(k)}$.

Obviously, the internal space is stabilized if potential \rf{5.3}
has a minimum with respect to $\ov \beta^{\, 1}$. The extremum
condition reads:
\be{5.4}
\left.\frac{\partial U_{int}}{\partial \ov \beta^{\, 1
}}\right|_{\ov \beta^{\, 1} =0} =0 \Longrightarrow
\frac{d_1D_0}{D_0-2}\lambda = \frac{2d_1}{D_0-2}\Lambda_D +
\frac{2d_1(D_0-1)}{D_0-2}\tilde f^{2}_1 \, .
\ee
The value of this potential at the minimum plays the role of
effective four-dimensional cosmological constant:
\be{5.5}
\Lambda_{eff} := \left. U_{int}\right|_{\ov \beta^{\, 1}=0} =
\Lambda_D + \tilde f^2_1 -\lambda  >0\, ,
\ee
which we assume to be positive. For the mass of gravexcitons we
obtain:
\be{5.6} m_{exci}^2 \sim \left.\frac{\partial^2 U_{int}}{\partial
{\ov \beta^{\, 1 }}^2}\right|_{\ov \beta^{\, 1} =0} =
\left(\frac{2d_1}{D_0-2}\right)^2\Lambda_D
+\left(\frac{2d_1(D_0-1)}{D_0-2}\right)^2\tilde f^{2}_1
-\left(\frac{d_1D_0}{D_0-2}\right)^2\lambda >0\, .
\ee
It can be easily seen from condition \rf{5.4} and inequality
\rf{5.5} that all three parameters are positive\footnote{Thus,
from $\lambda >0$ follows that summarized tensions of branes
should be negative.}: $\tilde f^{2}_1, \Lambda_D, \lambda
>0$. Taking into account inequality \rf{5.6}, it can be easily
verified that all these parameters have the same order of
magnitude:
\be{5.7}
\tilde f^{2}_1 \sim \Lambda_D \sim \lambda \sim \Lambda_{eff} \sim
m_{exci}^2\, .
\ee
Therefore, there is no need for fine tuning of parameters to
obtain the observable value of dark energy. To get it, it is
sufficient to suppose that all these parameters, including
$\Lambda_{eff}$, are of the order of $\Lambda_{DE} \sim
10^{-123}L_{Pl}$. Thus, we obtain the condition $\Lambda_{eff}\sim
\Lambda_{DE} \sim 10^{-57}\mbox{cm}^{-2}$. From one side, it is
natural to assume that parameters of the model have the same order
of magnitude. From other side, our model does not answer why this
value is equal to $10^{-123}L_{Pl}$. According to the anthropic
principle, it takes place because human life is possible only at
this value of dark energy.

If we assume that our parameters are defined by $\Lambda_{DE} \sim
10^{-123}L_{Pl}$, then we get for the gravexciton masses $m_{exci}
\sim 10^{-33}$eV $\sim 10^{-61}M_{Pl}$. These ultra-light
particles have a period of oscillations $t \sim 1/m_{\psi} \sim
10^{18}$sec which is of order of the Universe age. So, up to now
these cosmological gravexcitons did not start to oscillate but
slowly move to the position of minimum of the effective potential.
In this case it is hardly possible to speak about stabilization of
the internal space (the effective potential $U_{int}$ is too flat)
and we again arrive to the problem of the fundamental constant
variations.


\section{Conclusion}

\setcounter{equation}{0}

In the present paper we considered a possibility for the
construction of the conventional cosmology for our observable
Universe if underlying theory is multidimensional Kaluza-Klein
model. In the spirit of the Friedmann model, our multidimensional
models are also endowed with a perfect fluid as a matter source.
We investigated two different types of KK models.

For the first class of models, the internal spaces become
unobservable due to the dynamical compactification. Following an
ansatz proposed in Ref. \cite{Mohammedi}, we obtained effective
four-dimensional model which has the form of the Friedmann model
with the exception of two important things. First, the effective
four-dimensional gravitational "constant" $\kappa_0^2$ is a
dynamical function and, second, the energy density of effective
four-dimensional perfect fluid $\rho_{(4)}$ is not conserved (in
usual four-dimensional sense). Accordingly, the energy density
$\rho_{(4)}$ behaves differently than in the FRW Universe.
However, the combination $\kappa_0^2\rho_{(4)}$ has exactly the
same dependence on the scale factor as in the conventional
cosmology. Precisely this combination enters into the Friedmann
equations. Thus, the dynamical behavior of the external space in
this model exactly coincides with the standard Friedmann model,
i.e. here we have the same rate of the evolution. Unfortunately,
as our investigation shows, the effective four-dimensional
fundamental "constants" (e.g. the gravitational constant and the
fine structure constant) undergo too large variations with time in
comparison with the observations.

Then, we investigated the problem of the sable compactification of
the internal spaces. We proved the no-go theorem which claims that
the stable compactification of the internal spaces in KK models is
impossible in the general case with arbitrary equations of state
in the internal and external spaces. There are only two
exceptional classes with appropriate fitted equations of state.
{}From these classes of solutions, we constructed a model with the
standard behavior of the FRW Universe for the external space.
Moreover, this model is endowed with positive 4-D effective
cosmological constant. However, the fine tuning of parameters is
required to get the observable dark energy density. To avoid the
fine tuning problem, we generalized this model to the case of
orbifold internal space with branes in fixed points. In the spirit
of the Universal Extra Dimension models, the Standard Model fields
are not localized on the brane but can move in all bulk space. We
considered Ricci-flat orbifolds. The internal space stabilization
condition together with the positivity of the effective 4-D
cosmological constant $\Lambda_{eff}>0$ lead to the conclusion
that $\Lambda_{eff}$ can be of the order of observable dark energy
cosmological constant $\Lambda_{DE}$ without any fine tuning of
the parameters. However, in this case the effective potential is
too flat (mass gravexcitons is very small) to provide necessary
constancy of the effective fundamental "constants".



\mbox{} \\ {\bf Acknowledgments}\\ I thank the Organizing
Committee of the 14th International Seminar on High Energy Physics
"QUARKS-2006" in St. Petersburg for their financial support.


\end{document}